\documentclass[%
 reprint,
 superscriptaddress,
 amsmath,amssymb,
 aps,
]{revtex4-2}

\usepackage{graphicx}
\usepackage{dcolumn}
\usepackage{bm}

\begin{document}

\preprint{APS/123-QED}

\title{Quantum vortex phases of charged pion condensates induced by rotation in a magnetic field}

\author{Tao Guo}
\affiliation{School of Mathematics and Physics, Chengdu University of Technology, Chengdu 610059, China}

\author{Yanmei Xiao}
\affiliation{School of Mathematics and Physics, Chengdu University of Technology, Chengdu 610059, China}


\date{\today}%

\begin{abstract}

Using the relativistic complex scalar field model with a repulsive self-interaction, we discuss the ground state structure of charged pion condensation under the coexistence of parallel rotation and magnetic field.
Our previous study found that the density distribution profile of the condensates is a supergiant quantum vortex phase and change with rotational speed and coupling constant.
In this work, we further discover vortex lattice structures in the condensates under conditions of small rotation and strong coupling constant.
This mechanism can be thought of as electrical superconductivity: 
Vortex lattices are created to better adapt to changes in rotation and interaction.
Furthermore, large rotation and weak coupling constant are more likely to cause the vortex lattices to be destroyed and form a giant quantum vortex similar to a doughnut.
We expect this phenomenon can be observed in the relativistic non-central heavy ion collisions with large rotation and strong magnetic field.

\end{abstract}

\maketitle

\section{Introduction}

The realization of the Bose-Einstein condensate (BEC) with alkali metal elements provides physicists with a huge opportunity to study this new state of matter and marks the breakthrough development of modern physics. 
The applications of BEC involve theoretical and experimental research in many fields, such as ultracold atoms \cite{Anderson:1995gf,PhysRevLett.75.1687,PhysRevLett.123.160403}, superfluidity and quantum vortices \cite{PhysRevLett.83.2498,PhysRevLett.84.806,PhysRevLett.87.080402}. 
With in-depth research on BEC, physicists have extended traditional condensed matter physics to relativistic Bose-Einstein condensates (RBECs), i.e., the condensates are composed of relativistic microscopic constituents \cite{PhysRevLett.43.1277,PhysRevLett.99.200406,Fagnocchi_2010}.
As a consequence, in this era of rapid development of experimental facilities such as relativistic heavy ion collisions, it is of great physical significance to study the properties and forming mechanism of RBECs.

Rotation and magnetic field in general play very important roles in large variety of physical environments, such as rapidly rotating neutron stars \cite{Cook:1993qr,Bocquet:1995je,Andersson_1998}, binary black hole mergers \cite{Marronetti:2007wz,Pook-Kolb:2019iao} and non-central heavy ion collision experiments \cite{Becattini:2007sr,Becattini:2015ska,Jiang:2016woz}.
Especially, in non-central heavy ion collisions, the extreme conditions of strong magnetic fields and high rotational speeds have been realized \cite{Voronyuk:2011jd,Bzdak:2011yy,Kharzeev:2015znc}.
By analyzing experimental results, it can be concluded that the strong magnetic field generated during the collisions can be ${\rm e}B \sim m^2$ \cite{Tuchin:2013ie,STAR:2021mii}, where ${\rm e}$ is the value of an electron charge and $m$ is the mass of a charged pion.
In addition, the numerical simulations indicate that the angular momentum generated in the collisions can involve in the range [$10^3, 10^5$]~$\hbar$ \cite{Deng:2016gyh,Becattini:2020ngo};
and the rotational angular velocity has reached $\Omega \approx (9\pm1)\times10^{21}$ Hz $\sim$ 6.2 MeV \cite{STAR:2017ckg,STAR:2019erd}.
Generally, a strong external magnetic field can enhance a fermion-antifermion condensation for leading to generate a fermion dynamical mass \cite{Klimenko:1991he,PhysRevLett.73.3499,Miransky:2015ava}.
Research shows that the rotation has similar effect to a magnetic field and can cause certain anomalous transport processes, for example, chiral vortical
effect \cite{Kharzeev:2007tn,Kharzeev:2010gr} and chiral vortical wave \cite{Jiang:2015cva,Kamada:2022nyt}.
While, opposite to the magnetic catalysis effect, the rotation generally suppresses the chiral condensation at finite temperature according to effective models \cite{Jiang:2016wvv,Ebihara:2016fwa,Chen:2022mhf}. 
However, the lattice QCD simulations at imaginary rotation seem to deny the latter feature, which then cause a lot of debate and discussion on rotation effect \cite{Braguta:2021jgn,Yang:2023vsw,Cao:2023olg}.

Recently, many studies have shown that the combination of parallel rotation and magnetic field (PRM) can also induce a variety of condensed distributions.
For instance, based on the solutions of the Dirac equation, the condensed characteristics of free fermionic systems in PRM have been discussed \cite{PhysRevD.26.1900,Chen:2015hfc,Liu:2017zhl,Fukushima:2020ncb}.  
Also, for interacting rotating fermion systems in a magnetic field, more research focuses on the possible effects of the edge states in phase structure \cite{McInnes:2016dwk,Hattori:2016njk,Chernodub:2016kxh,Fukushima:2018grm}.
Within the three-flavor Nambu–Jona-Lasinio (NJL) model, the PRM can induce charged rho ($\rho^{\pm}$) superconductor even at a small rotational angular velocity \cite{Cao:2020pmm}.
The combined effects of PRM can make non-interacting charged pions condense both in the vacuum and at finite temperature \cite{PhysRevLett.120.032001}.
However, other calculations show that the charged pion condensates can only occur under the conditions of a strong coupling constant and negatively large baryon chemical potential [\onlinecite{Chen:2019tcp}].
In our previous study \cite{Guo:2021gbz}, based on the viewpoint of spontaneous symmetry breaking, the results show that the profile of ground state formed by interacting charged pions in PRM is a supergiant quantum vortex.
Such possibility was verified by applying the Ginzburg-Landau analysis in the NJL model \cite{Cao:2019ctl}.

This work extends the previous one \cite{Guo:2021gbz} by looking into a wider range of interaction strength, espacially, the strong case.
This paper is organized as follows.
In Sec. \ref{S2}, we introduce the relativistic complex scalar field model with a repulsive self-interaction.
In Sec. \ref{S3}, we obtain the energy dispersion relation and wave function of free charged pion fields by solving the Klein-Gordon equation, and calculate the ground state density distribution of interacting charged pion fields induced by PRM.
We summarize in Sec. \ref{S4}.
The nature units $c = \hbar = k_B = 1$ are used throughout.

\section{Relativistic complex scalar field model \label{S2}}

In order to explore the ground state formation of the charged pion condensates induced by PRM, we adopt the relativistic complex scalar field model with a repulsive self-interaction.
In the cylindrical coordinates $\left (r, \theta, z\right )$, the action of the relativistic complex scalar field can be given in a curved spacetime by
\begin{equation}
	{\cal S} = \int dt \int d^3 \textbf{x} \sqrt{- {\rm det} (g_{\mu\nu})} {\cal L}(\Phi^\ast, \Phi),
\end{equation}
where $d^3 \textbf{x} = r dr d\theta dz $, and the Lagrangian density is defined by
\begin{equation}
	{\cal L} = \left(D^\mu \Phi\right)^*\left(D_\mu\Phi\right) -m^2|\Phi|^2  - g |\Phi|^4.
\end{equation}
Here the covariant derivative $D_\mu = \partial_\mu +i{\rm e}A_\mu$, where $A_\mu$ is the vector potential. 
$\Phi$ represents a charged complex scalar field. 
$g$ is a coupling constant that reflects the strength of the self-interaction.
The spacetime metric $g_{\mu\nu}$ of the rotating frame reads
\begin{equation}
g_{\mu\nu}=\left(
\begin{array}{cccc}
1-r^2\Omega^2 & y\Omega & -x\Omega & 0\\
y\Omega           & -1 & 0 & 0\\
-x\Omega    & 0 & -1 & 0 \\
0           & 0 &  0 & -1
\end{array}\right),
\end{equation}
where $r=\sqrt{x^2+y^2}$ is the radius of the cylinder system, the notation $\Omega$ is the rotational angular velocity, and $\sqrt{- {\rm det} (g_{\mu\nu})} = 1$.
It is convenient to rewrite the Lagrangian density of the system as
\begin{eqnarray}\nonumber
	\mathcal{L} = &&|\left (D_t+\Omega y D_x - \Omega x D_y\right )\Phi|^2 - |D_i\Phi|^2\\ 
	  &&- m^2|\Phi|^2 - g |\Phi|^4.
\end{eqnarray}
Here the Lagrangian density is obviously invariant under the local $U(1)$ symmetry. 
Moreover, physical properties are generally not influenced by the specific choice of gauge.
For convenience, we choose the symmetrical gauge in the following calculations.

In the rotating frame, the Lagrangian density for the interacting charged pion fields in the symmetric gauge can be expressed as
\begin{equation}\label{Lagran}
	\mathcal{L} = |(\partial_t-i\Omega L_z)\Phi|^2 - |D_i\Phi|^2 -m^2|\Phi|^2  - g |\Phi|^4,
\end{equation}
where $L_z = -i\partial_\theta = -i(x\partial_y - y\partial_x)$ is the angular momentum along the $z$-axis. In the imaginary time formalism, the partition function of the system is given by
\begin{equation}\label{pf}
	\mathcal{Z} = \int \left[d\Phi\right]\left[d\Phi^\ast\right] {\rm exp} \left (\int_0^\beta d\tau \int d^3 \textbf{x} \mathcal{L}\right ).
\end{equation}
Here $\beta$ ($= 1/T$) is defined as the inverse temperature, and $\tau$ is the imaginary time. 
By integrating the above partition function (\ref{pf}) by parts, we can rewrite the action of the system as
\begin{equation}\label{action}
	{\cal S} = -\int_0^\beta d\tau \int d^3 \textbf{x} \mathcal{L} = \int_0^\beta d\tau \int d^3 \textbf{x} \mathcal{H},
\end{equation}
where the notation ${\cal H}$ reads
\begin{eqnarray}\nonumber
	{\cal H} &=& \Phi^\ast\left[-\left (\partial_\tau -\Omega L_z\right )^2\right]\Phi \\\nonumber
	&+& \Phi^\ast\left(- \Delta + \frac{1}{4}{\rm e}^2B^2r^2 - {\rm e}BL_z + m^2\right)\Phi \\
	&+& g |\Phi|^4.
\end{eqnarray}
The operator $\Delta$ represents the Laplace operator in the cylindrical coordinate frame
\begin{equation}
	\Delta \equiv \nabla^2 = \partial_r^2+\frac{1}{r}\partial_r - \frac{L_z^2}{r^2}  + \partial_z^2.
\end{equation}

\section{Ground state formation of charged pion condensates \label{S3}}

If the ground state density distribution of the system is a Bose-Einstein condensate, the charged pion field $\Phi$ acquires a nonzero expectation value.
Thus we can decompose the charged pion field $\Phi$  into a classical part $\phi_0(\textbf{x})$ and a quantum fluctuation part $\phi(\tau, \textbf{x})$, i.e.,
\begin{equation}
	\Phi = \phi_0\left( \textbf{x} \right) +\phi \left(\tau,\textbf{x}\right).
\end{equation}
We note that in a finite-size system, the condensate $\phi_0(\mathbf{x})$ is generally inhomogeneous. 
At zero temperature, the density distribution profile of the charged pion condensates can be determined by minimizing the Gross-Pitaevskii-like free energy 
\begin{eqnarray}\label{gpl-fe}\nonumber
		E &=& \int d^3 \textbf{x} \left[\phi_0^\ast \left( \textbf{x} \right) \left(  -\Delta + H \right) \phi_0\left(\textbf{x} \right) \right] \\
	 	  &+& g \int d^3 \textbf{x} \left|\phi_0\left( \textbf{x} \right)\right|^4,
\end{eqnarray}
where the operator $H$ is defined as
\begin{equation}
	H = \frac{1}{4}{\rm e}^2B^2r^2 - eBL_z - \Omega^2L_z^2 + m^2.
\end{equation}
In principle, considering that the condensate in the $z$-axis direction is homogeneous, we can simplify the Gross-Pitaevskii-like free energy as
\begin{eqnarray}\label{gpl-fe}\nonumber
	K &=& \frac{E}{\int dz}\\\nonumber
	&=&\int d \textbf{r} \left[\phi_0^\ast \left( \textbf{r} \right) \left( - \partial_r^2 - \frac{1}{r}\partial_r + \frac{ L_z^2}{r^2} + H \right) \phi_0\left(\textbf{r} \right) \right] \\
          	 &+& g \int d \textbf{r} \left|\phi_0\left( \textbf{r} \right)\right|^4,
\end{eqnarray}
where $\int d \textbf{r} = \int_0^{2\pi} d \theta \int_0^R r dr$ with $\textbf{r} \equiv (r, \theta)$.
And $R$ is the radius of the cylinder cross section.

In the following, we define a certain number of conserved charges.
The ground state features of free and interacting charged pion fields are studied separately.

\subsection{Klein-Gordon equation of free charged pion fields}
 
Let us first solve the more general Klein-Gordon equation of free charged pion fields.
We consider the charged pion fields in the cylindrical coordinate with a uniform magnetic field $\boldsymbol{B}=B\vec{z}$ and a constant rotation $\boldsymbol{ \Omega}=\Omega\vec{z}$. 
In this paper, we always take ${\rm e} B > 0$ and $\Omega > 0$, unless otherwise stated.
The Klein-Gordon equation of free charged pion fields in a rotating frame can be given by
\begin{equation}\label{KGE1}
	- \left (\partial_t- i \Omega L_z \right )^2\Phi + D_i^\dagger D_i\Phi -m^2\Phi = 0,
\end{equation}
with the derivative $D_t = \partial_t -i{\rm e}B\Omega r^2/2$, $D_x = \partial_x +i{\rm e}By/2$, $D_y = \partial_y -i{\rm e}Bx/2$, $D_z = \partial_z$. 
Therefore, the solution for the above equation (\ref{KGE1}) can be written as
\begin{equation}\label{solution}
	\Phi = e^{-i\varepsilon t + i p_z z}\phi_{nl}\left( \textbf{r} \right),
\end{equation}
with  
\begin{equation}\label{wfk}
	\phi_{nl}\left( \textbf{r} \right) = {\cal C} e^{ il\theta}\phi_{nl}\left( r \right).
\end{equation}
The notation $p_z$ is the momentum along the $z$-direction, ${\cal C}$ is the normalization factor, and $l$ is the azimuthal angular quantum number.

In the infinite-size volume case, substituting (\ref{solution}) to (\ref{KGE1}), we can rewrite the Klein-Gordon equation
\begin{equation}\label{KGE2}
 \left[\left(\varepsilon +  \Omega L_z \right)^2 + \Delta  - H \right]\phi_{nl}\left( \textbf{r} \right) = 0.
\end{equation}
Due to the need to guarantee the causal conditions of relativity, we must consider the rotation with a finite velocity $v = \Omega  r \le 1$ and the quantization condition justify with $r \gg 1/\sqrt{{\rm e}B}$. 
When the first inequality is imposed, the frame does not move faster than light to avoid pathological effects of particle spectrum. 
The second inequality is satisfied for keeping the wave function localization at the boundary. 
In this way, the interference of some non-physical factors can be eliminated, so that the real physical results can be better obtained.
In general, we assume that the rotation is rigid, so the rotational angular velocity $\Omega$ does not depend on the distance to the axis-of-rotation.

In the above case, it is convenient for obtaining the radial solution of the equation (\ref{KGE2})
\begin{equation}
	\phi_{nl}\left (r\right )  =  r^{|l|} e^{-\frac{1}{4}{\rm e}Br^2}{_1F_1}\left (-a_{nl},|l|+1,\frac{{\rm e}Br^2}{2}\right ),
\end{equation}
where ${_1F_1}$ is a confluent hypergeometrical function with the parameter
\begin{equation}\label{anl}
	a_{nl} = \frac{1}{2{\rm e}B} \left[\left (\varepsilon + \Omega l\right )^2 - p_z^2 - m^2\right] - \frac{1}{2}\left (|l| - l +1\right ).
\end{equation}
Therefore, we can derive the expression of the energy dispersion relationship from (\ref{anl}) as
\begin{equation}
	\left (\varepsilon+\Omega l\right )^2 = p_z^2 + m^2 + {\rm e}B\left (2a_{nl}+|l|-l+1\right ).
\end{equation}
Here the parameter $a_{nl}$ is the $n$-th zero point value of the confluent hypergeometric function ${_1F_1}$ when the angular quantum number $l$ is given, 
i.e., $a_{nl}$ can be obtained by solving the equation
\begin{equation}
	{_1F_1}\left (-a_{nl},|l|+1,\frac{{\rm e}Br^2}{2}\right )=0.
\end{equation}
Especially, if we consider an infinite system with a radius $r \rightarrow \infty$, $a_{nl}$ is a set of non-negative integers $n$. 
The function ${_1F_1}$ can safely be simplified to an associated Laguerre polynomial.
Considering the non-rotation condition again, the energy dispersion relation of the system returns to the Landau levels (LL): $\varepsilon^2 = p_z^2 + m^2 +{\rm e}B \left (2n + 1\right )$. 
It is obvious that the LL with different $l$ is degenerate.
However, when a rotational angular velocity $\Omega$ is imposed on the system, the LL produces a shift $\mp \Omega l$.
Here, the angular quantum number of the positive charged particles is $l$ and the angular quantum number of the negative charged particles is $-l$.
As a result, this means that the positive charged particles split downward and the negative charged particles split upward.

In a real physical system, we choose the cylinder with cross-sectional radius $ r=R $.  
And we impose the Dirichlet boundary condition that ensures that the wave function of the charged pion fields must vanish at the edge of the cylinder. 
In the disc plane transverse to the rotating axis $\vec{z}$, each momentum $p_z$ corresponds to a Landau degeneracy factor
\begin{equation}
	N_f = \left[\frac{{\rm e}BS}{2\pi}\right]=\left[\frac{{\rm e}BR^2}{2}\right],
\end{equation}
where the square bracket [$\cdots$] means a rounding function.
For the LL to fit into the cross section disc with the area $S=\pi R^2$, the degeneracies of the LL are identified with the $z$-component of the angular momentum in position space. 
As a consequence, the possible range of the $l$ should be $-n \le l \le N_f -n$ where $n$ labels the LL.
And when $n=0$, it means the lowest Landau level (LLL) with $0 \le l \le N_f$.

\begin{figure*}[t]
	\centering
	\includegraphics[width=1.0\textwidth,height=0.79\textwidth]{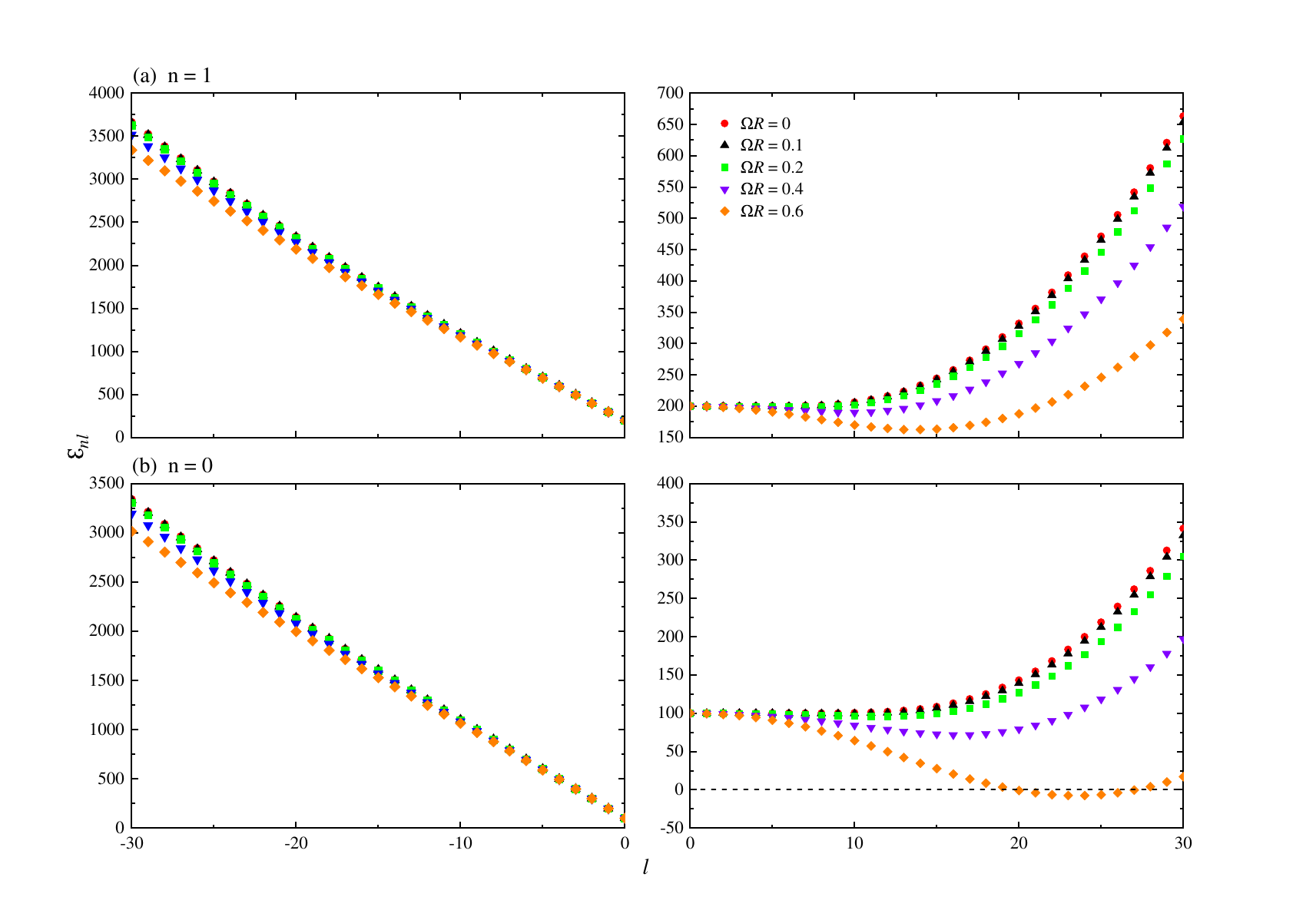}
	\vspace{-1cm}
	\caption{\label{f1} 
		Energy spectra of the lowest two energy levels ($n=0$ and $n=1$ ) as a function of the angular quantum number $l$ and different rotational speed $\Omega R$. 
		In this figure we take $N_f = 25$ and $R=10$ fm.}
\end{figure*}

Now we turn the problem to the free charged pion fields (i.e., $g = 0$) in the coexistence of PRM.
Compared with the solution of the corresponding Klein-Gordon equation, the global minimum of the Gross-Pitaevskii-like free energy of the system can be obtained by solving for the global minimum of $K_{nl}$, where the $K_{nl}$ reads
\begin{equation}\label{fKnl}
	K_{nl} =   - \Omega^2l^2 + {\rm e}B \left (2a_{nl} + |l|-l + 1\right ) + m^2.
\end{equation}
In order to conveniently consider the relativistic causality, the above (\ref{fKnl}) is rewritten as
\begin{equation}\label{fKnl1}
	{\cal E}_{nl} =   -  \left( \Omega R \right)^2l^2 + 2 N_f \left (2a_{nl} + |l|-l + 1\right ) + m_0 ^2,
\end{equation}
where ${\cal{E}}_{nl} = R^2 K_{nl}$ and $m_0 = R^2 m$. 
In this paper, we take the uniform magnetic field ${\rm e}B = m^2 \approx 0.5$ fm$^{-2}$.
Through the above results, we plot the two lower energy spectra of the free charged pion fields in (\ref{fKnl1}) as FIG. \ref{f1}. 
The behavior of other higher energy levels is similar.

It can be clearly concluded from FIG. \ref{f1} that the lowest energy of the $l <0 $ is much higher than the lowest energy of the $l> 0 $.
And with the increase of rotational angular velocity $\Omega$, the ground state of the free charged pion system is more inclined to the position with positive larger $l$. 
Obviously, this shows that the charged pion condensates of the positive $l$ modes are more favorable than the negative $l$ modes, i.e., the $l \ge 0$ modes always correspond to the ground state of the system.
Specifically, the $l$ corresponding to the ground state is the angular quantum number at the global minimum of $K_{nl}$ (or ${\cal{E}}_{nl}$). 
For example, when the rotational speeds $\Omega R$ are $0$, $0.1$, $0.2$, $0.4$ and $0.6$, the locations of the $l$ corresponding to the ground state are $0$, $9$, $12$, $16$ and $24$, respectively.
By considering (\ref{fKnl}), we can get that the global minimum of the system depends on the competition between $-\Omega^2l^2$ and $2{\rm e}B \left (2a_{nl} + 1\right ) + m^2$.
This indicates that $\Omega l$ in the rotating system plays the role of an effective $l$-dependent chemical potential.

\begin{figure*}[t]
	\centering
	\includegraphics[width=1\textwidth,height=0.83\textwidth]{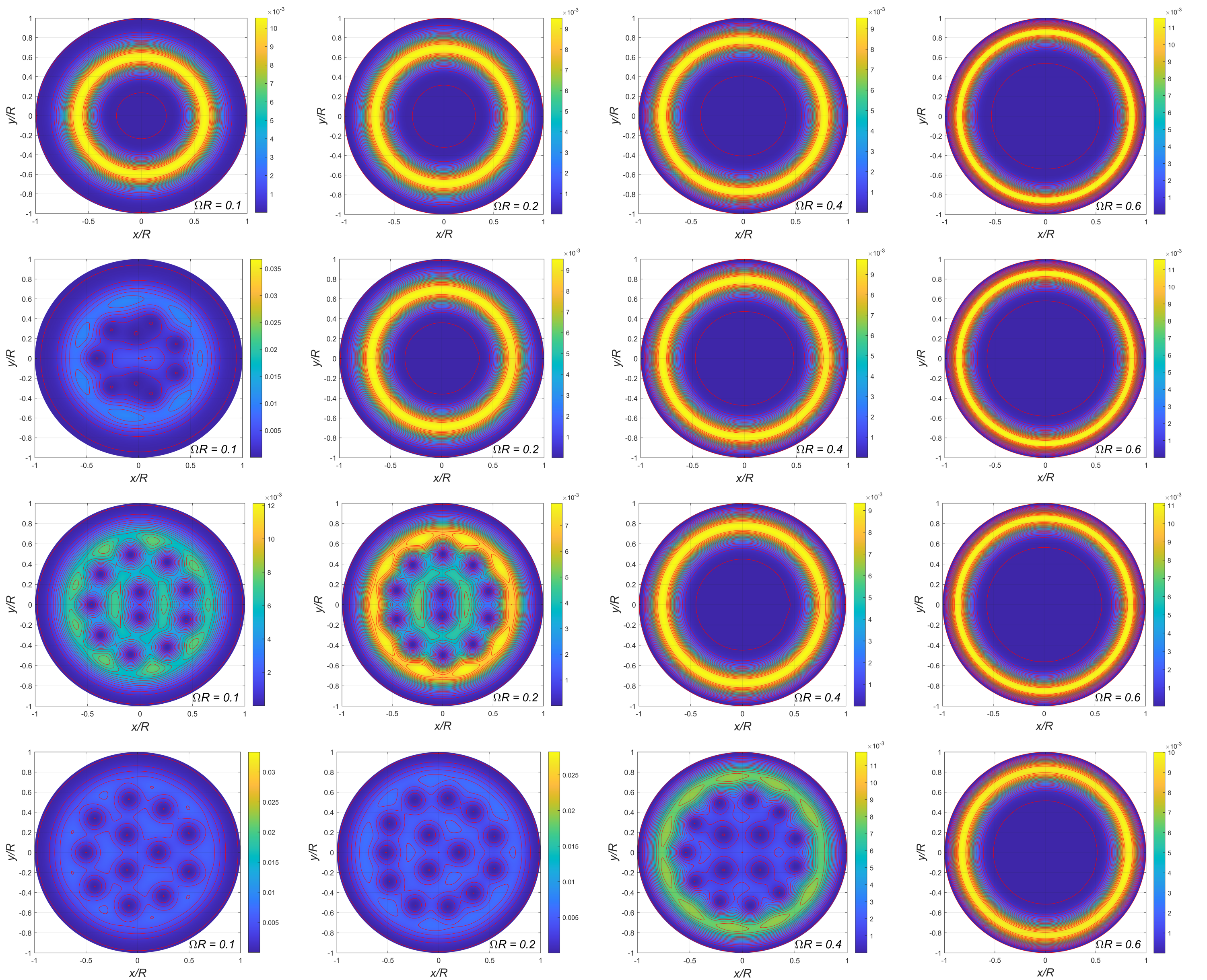}
	\caption{\label{f2} 
		In the coexistence of PRM, the ground state density distribution $|\phi_0\left( \textbf{r} \right)|^2$ of interacting charged pion fields changes with different rotational speeds $\Omega R $ and coupling constants $ g $. 
		Each column has the same $\Omega R$, and each row has the same $g$.
		The rotational speeds of different columns increase from left to right and are $\Omega R = 0.1, 0.2, 0.4, 0.6$, respectively.
		The coupling constants of different rows increase from top to bottom and are $g =0, 0.01, 0.1, 0.3$, respectively.
		In this figure we take $N_f = 25$ and $R=10$ fm.
	}
\end{figure*}

\subsection{Quantum vortex phases of interacting charged pion fields}
At zero temperature, the minimization of the Gross-Pitaevskii-like free energy (\ref{gpl-fe}) leads to the ground state energy spectra of interacting charged pion fields.
The equation of motion of interacting charged pion system, however, is nonlinear and cannot be solved analytically.
Therefore, we choose the variational method to numerically calculate the lowest energy eigenstate of the system.
Considering the basic features of interacting charged pion fields, the trial wavefunction $\phi_0\left( \textbf{r} \right)$ can be expressed by the complete basis vectors composed of the wave functions of free charged pion fields as 
\begin{equation}\label{wf}
	\phi_0\left( \textbf{r} \right) = \sum_{n=0}^{\infty}\sum_{l=-\infty}^{\infty} c_{nl}  \phi_{nl}\left( \textbf{r} \right).
\end{equation}
The variational parameters $c_{nl}$ are determined by minimizing the Gross-Pitaevskii-like free energy.
For a given number of charged pions, the conserved charges $\delta N$ is defined as
\begin{equation}\label{cc}
	\delta N  \equiv \sum _ {nl} \left(N_{nl}  - \overline{N}_{nl}\right) = \sum_{nl} |c_{nl}|^2,
\end{equation}
where $N_{nl} \left( \overline{N}_{nl} \right)$ is the number of positive pions (negative pions).
In realistic numerical calculations, we can equally choose the probability density $\sum_{nl} \left| c_{nl} \right|^2 = 1$.
Substituting the trial wavefunction (\ref{wf}) into (\ref{gpl-fe}), this Gross-Pitaevskii-like free energy can be written as the sum of two parts: $K = K_0 + K_{int}$.
Here we take the non-interacting part $K_0 = \sum_{n=0}^{\infty}\sum_{l=-\infty}^{\infty} |c_{nl}|^2 \varepsilon_{nl}$ and the interacting part $K_{int} = g \int d \mathbf{r} |\phi_0(\mathbf{r})|^4$.  
According to the dispersion relationship (\ref{fKnl}), we can obtain $\varepsilon_{nl} = - \Omega^2l^2 + {\rm e}B (2a_{nl} + 1) + m^2$.
Obviously, $K$ is the superposition of the quadratic and quartic term.
The minimum of the system still depends mainly on the competition between $- \Omega^2l^2$ and ${\rm e}B (2a_{nl} + 1) + m^2$.

By doing complicated numerical solution, we get the global minimum (i.e., the ground state energy of interacting charged pions) of the Gross-Pitaevskii-like free energy (\ref{gpl-fe})  and the corresponding ground state $\phi_0\left( \textbf{r} \right)$.
Without loss of generality, we take four different rotational speeds $\Omega R$ and four different coupling constants $g$.
The ground state phase diagrams of the interacting charged pion condensates are shown in FIG. \ref{f2}.
It should be noted that we did not plot ground state density distribution for the $\Omega = 0$ here.
Because when $\Omega =0$ and $g = 0$, the interacting charged pion fields are restored to the free non-rotational charged pion fields.
Profile of the condensate shows that almost all charged pions condense in the center (i.e., the state with $l = 0$) of the disc.
This phenomenon is similar to the traditional BEC, in which all bosons occupy the same quantum state to form a macroscopic observable.
When $\Omega =0$ and $g \neq 0$, the ground state density distribution is diffuse in the two-dimensional plane.
In FIG. \ref{f2}, it can be found that the profiles of these charged pion ground states are various quantum vortex phases.
And some of these condensates are characterized by the creation of a certain number of vortex lattices.
These vortex lattices are similar to small twisters inside the flowing liquids.

In the absence of interaction ($g = 0$, see the first row in FIG. \ref{f2}), the profile of the condensates will gradually expand from the center to the edge position with increasing of the $\Omega R$.
However, it is worth noting that this expansion change is not continuous, but quantized.
Besides, pure rotation cannot effectively induce multiple vortex lattices, but makes the charged pions condense into a specific single particle state with a determined $l$.
As previously analyzed, the rotation causes a shift in the ground state energy level.
Specifically, the ground state changes from  $\varepsilon$ to $\varepsilon - \Omega l$ corresponding to the angular quantum number from $l=0$ to $l \neq 0$, respectively.
Therefore, the profile of the charged pion condensates is actually a single quantum vortex state when $\Omega R \neq 0$ and $g = 0$. 
The radius of the single quantum vortex depends on the rotational angular velocity. 
Quantum vortices generally have unique intrinsic angular momentum that is different from spin.
In a cylinder, the wave function of the quantum vortex states can be written as
\begin{equation}\label{vwf}
	\phi_0\left( \textbf{r} \right) = {\eta} \left(r\right) e^{il_c\theta},
\end{equation}
where $\eta \left(r\right) $ is just a $r$-dependent function, and $\eta \left(r\right) \propto r^{|l_c|} $.
$l_c$ is a non-negative integer and is also defined as the winding number (or topological charge) in the vortex state.
Theoretically, $l_c$ is proportional to the angular momentum quantum number.
From the first row in FIG. \ref{f2}, we can find that the larger $\Omega R$, the larger $l_c$.
Numerical calculations show that $l_c = l$, when the rotating charged pions are in the ground state.
In particular, the results show that rotation has a certain catalytic effect in a magnetic field.
This allows the non-interacting charged pions to form a single giant quantum vortex with a winding number $l_c \gg 1$ induced by PRM.

In the following, we turn our attention to the system of interacting charged pion fields ($g \neq 0$, see the rows 2-4 in FIG. \ref{f2}).
We find that the formation of vortex lattices results from conditions of small rotation and strong coupling constant.
In a uniform magnetic field, the possible reason for the formation of vortex lattices is to trigger electrical superconductivity to adapt to the combined effects of interaction and rotation.
The number of vortex lattices reflects the weight of free charged pion states with different $l$.
It can generally be represented by the values of the non-vanishing parameters $c_{nl}$ in (\ref{wf}).
For example, when $g = 0.1$ and $\Omega R = 0.1$ (i.e., the subfigure in the third row and first column in FIG. \ref{f2}), there are two rings of vortex lattices in the ground state.
The non-vanishing parameters $c_{nl}$ are $c_{0l} = (0.144, 0.516, 0.835)$ for $l = (0, 2, 11)$, respectively.
In this quantum vortex structure, it is shown that there are two vortex lattices in the inner ring, nine vortex lattices in the outer ring and the total number is eleven.
This result accurately shows that the maximum $l$ contained in the ground state is the total number of vortex lattices.
In particular, the features of other subfigures in FIG. \ref{f2} are similar.

Large rotation and weak coupling constant are more likely to cause the vortex lattices to be destroyed and form a new single giant quantum vortex.
As the $\Omega R$ continues to increase, more charged pions are condensed on the edge of the disc to form a doughnut-like structure.
This change demonstrates that the larger rotation destroys the equilibrium state initially formed in the condensates, prompting the transformation of multiple vortex lattices into a single giant quantum vortex state.
The reason for this transformation process is probably the effect of centrifugal force caused by large rotation.
And the single giant quantum vortex state has a certain winding number $l_c$.
For example, when $\Omega R = 0.6$ (see the fourth column in FIG. \ref{f2}), the winding numbers are $l_c = 24, 24, 24, 23$ for $g =0, 0.01, 0.1, 0.3$, respectively. 
Here, the small difference in the $l_c$ results from the coupling of adjacent energy levels due to the existence of an interaction.
At this point, the interaction effect is significantly smaller compared to rotation.
Obviously, this suggests that rotation and interaction form an entanglement effect in the charged pion condensates.

The profile of the condensates may be even richer if we consider dynamic electromagnetic fields. 
As we all know, the rotation and magnetic field have similar effects. 
And the effect of rotation applies not only to charged particles, but also to neutral particles.
For free charged pion fields, the rotation causes a change in the ground state of the system from $l=0$ to $l=l_c$.
This change is generally expressed as an effective $l$-dependent chemical potential.
For interacting charged pion fields, the combined effect of rotation and magnetic field may produce more meaningful ground state density distribution with the various vortex lattice structures.
These provide us with a deeper understanding of the evolution of condensates induced by PRM.
Of course, some signals of QGP and CME can also be studied through the properties of vortex phase formed in the condensates.
Therefore, it is of principal interest that we look forward to more quantitative studies in the next work.

\section{summary \label{S4}}

In PRM, we calculate the ground state density distribution of charged pions formed under a wider range of interactions and rotational speeds.
The method is based on the relativistic complex scalar field model with a repulsive self-interaction and then solved by variational calculations.
We conclude that a certain number of vortex lattices are found in the condensates under conditions of small rotation and strong coupling constant.

Our calculation results show that for free charged pion fields, the condensate always tends to angular quantum number $l \ge 0$ modes. 
And as the rotational angular velocity $\Omega$ increases, the corresponding $l$ of the ground state increases.
This indicates that $\Omega l$ in the rotating system plays the role of an effective $l$-dependent chemical potential.
For interacting charged pion fields, we find that as coupling constant increases, the vortex lattices are generated in the ground state of charged pions.
In this case, this mechanism can be thought of as electrical superconductivity. 
These vortex lattices are created to better adapt to changes in rotation and interaction.
Under conditions of large rotation and weak coupling constant, the vortex lattices inside disc tends to be destroyed and more pions condense on the edge of the disc to form a doughnut-like structure.
This phenomenon is similar to the condensate of free charged pions in a finite-size cylinder system with a large rotation, and the reason can be considered to be the result of centrifugal force.
Moreover, these vortex structures are theoretically observable.

Finally, a natural extension of the present paper will be a more self-consistent and realistic research of the profile of charged pion condensates (i.e., the condensed distribution under dynamic electromagnetic fields,  the pion superfluid in non-central heavy-ion collisions), which may lead to more discussions about the features of the RBECs in a finite-size rotating system.
We hope the measurement of multipion correlations can help us to check the formation and evolution of quantum vortex phases in the charged pion condensates under parallel magnetic field and rotation.

\begin{acknowledgments}
The authors would like to thank Fan Wu for helpful comments and discussions. This work was supported by the Scientific Research Foundation of Chengdu University of Technology under grant No.10912-KYQD2022-09557.

\end{acknowledgments}


\bibliography{vsref}

\providecommand{\noopsort}[1]{}\providecommand{\singleletter}[1]{#1}%
\begin{thebibliography}{52}%
\makeatletter
\providecommand \@ifxundefined [1]{%
 \@ifx{#1\undefined}
}%
\providecommand \@ifnum [1]{%
 \ifnum #1\expandafter \@firstoftwo
 \else \expandafter \@secondoftwo
 \fi
}%
\providecommand \@ifx [1]{%
 \ifx #1\expandafter \@firstoftwo
 \else \expandafter \@secondoftwo
 \fi
}%
\providecommand \natexlab [1]{#1}%
\providecommand \enquote  [1]{``#1''}%
\providecommand \bibnamefont  [1]{#1}%
\providecommand \bibfnamefont [1]{#1}%
\providecommand \citenamefont [1]{#1}%
\providecommand \href@noop [0]{\@secondoftwo}%
\providecommand \href [0]{\begingroup \@sanitize@url \@href}%
\providecommand \@href[1]{\@@startlink{#1}\@@href}%
\providecommand \@@href[1]{\endgroup#1\@@endlink}%
\providecommand \@sanitize@url [0]{\catcode `\\12\catcode `\$12\catcode
  `\&12\catcode `\#12\catcode `\^12\catcode `\_12\catcode `\%12\relax}%
\providecommand \@@startlink[1]{}%
\providecommand \@@endlink[0]{}%
\providecommand \url  [0]{\begingroup\@sanitize@url \@url }%
\providecommand \@url [1]{\endgroup\@href {#1}{\urlprefix }}%
\providecommand \urlprefix  [0]{URL }%
\providecommand \Eprint [0]{\href }%
\providecommand \doibase [0]{https://doi.org/}%
\providecommand \selectlanguage [0]{\@gobble}%
\providecommand \bibinfo  [0]{\@secondoftwo}%
\providecommand \bibfield  [0]{\@secondoftwo}%
\providecommand \translation [1]{[#1]}%
\providecommand \BibitemOpen [0]{}%
\providecommand \bibitemStop [0]{}%
\providecommand \bibitemNoStop [0]{.\EOS\space}%
\providecommand \EOS [0]{\spacefactor3000\relax}%
\providecommand \BibitemShut  [1]{\csname bibitem#1\endcsname}%
\let\auto@bib@innerbib\@empty
\bibitem [{\citenamefont {Anderson}\ \emph {et~al.}(1995)\citenamefont
  {Anderson}, \citenamefont {Ensher}, \citenamefont {Matthews}, \citenamefont
  {Wieman},\ and\ \citenamefont {Cornell}}]{Anderson:1995gf}%
  \BibitemOpen
  \bibfield  {author} {\bibinfo {author} {\bibfnamefont {M.}~\bibnamefont
  {Anderson}}, \bibinfo {author} {\bibfnamefont {J.}~\bibnamefont {Ensher}},
  \bibinfo {author} {\bibfnamefont {M.}~\bibnamefont {Matthews}}, \bibinfo
  {author} {\bibfnamefont {C.}~\bibnamefont {Wieman}},\ and\ \bibinfo {author}
  {\bibfnamefont {E.}~\bibnamefont {Cornell}},\ }\bibfield  {title} {\bibinfo
  {title} {{Observation of Bose-Einstein condensation in a dilute atomic
  vapor}},\ }\href {https://doi.org/10.1126/science.269.5221.198} {\bibfield
  {journal} {\bibinfo  {journal} {Science}\ }\textbf {\bibinfo {volume}
  {269}},\ \bibinfo {pages} {198} (\bibinfo {year} {1995})}\BibitemShut
  {NoStop}%
\bibitem [{\citenamefont {Bradley}\ \emph {et~al.}(1995)\citenamefont
  {Bradley}, \citenamefont {Sackett}, \citenamefont {Tollett},\ and\
  \citenamefont {Hulet}}]{PhysRevLett.75.1687}%
  \BibitemOpen
  \bibfield  {author} {\bibinfo {author} {\bibfnamefont {C.~C.}\ \bibnamefont
  {Bradley}}, \bibinfo {author} {\bibfnamefont {C.~A.}\ \bibnamefont
  {Sackett}}, \bibinfo {author} {\bibfnamefont {J.~J.}\ \bibnamefont
  {Tollett}},\ and\ \bibinfo {author} {\bibfnamefont {R.~G.}\ \bibnamefont
  {Hulet}},\ }\bibfield  {title} {\bibinfo {title} {Evidence of bose-einstein
  condensation in an atomic gas with attractive interactions},\ }\href
  {https://doi.org/10.1103/PhysRevLett.75.1687} {\bibfield  {journal} {\bibinfo
   {journal} {Phys. Rev. Lett.}\ }\textbf {\bibinfo {volume} {75}},\ \bibinfo
  {pages} {1687} (\bibinfo {year} {1995})}\BibitemShut {NoStop}%
\bibitem [{\citenamefont {Tononi}\ and\ \citenamefont
  {Salasnich}(2019)}]{PhysRevLett.123.160403}%
  \BibitemOpen
  \bibfield  {author} {\bibinfo {author} {\bibfnamefont {A.}~\bibnamefont
  {Tononi}}\ and\ \bibinfo {author} {\bibfnamefont {L.}~\bibnamefont
  {Salasnich}},\ }\bibfield  {title} {\bibinfo {title} {Bose-einstein
  condensation on the surface of a sphere},\ }\href
  {https://doi.org/10.1103/PhysRevLett.123.160403} {\bibfield  {journal}
  {\bibinfo  {journal} {Phys. Rev. Lett.}\ }\textbf {\bibinfo {volume} {123}},\
  \bibinfo {pages} {160403} (\bibinfo {year} {2019})}\BibitemShut {NoStop}%
\bibitem [{\citenamefont {Matthews}\ \emph {et~al.}(1999)\citenamefont
  {Matthews}, \citenamefont {Anderson}, \citenamefont {Haljan}, \citenamefont
  {Hall}, \citenamefont {Wieman},\ and\ \citenamefont
  {Cornell}}]{PhysRevLett.83.2498}%
  \BibitemOpen
  \bibfield  {author} {\bibinfo {author} {\bibfnamefont {M.~R.}\ \bibnamefont
  {Matthews}}, \bibinfo {author} {\bibfnamefont {B.~P.}\ \bibnamefont
  {Anderson}}, \bibinfo {author} {\bibfnamefont {P.~C.}\ \bibnamefont
  {Haljan}}, \bibinfo {author} {\bibfnamefont {D.~S.}\ \bibnamefont {Hall}},
  \bibinfo {author} {\bibfnamefont {C.~E.}\ \bibnamefont {Wieman}},\ and\
  \bibinfo {author} {\bibfnamefont {E.~A.}\ \bibnamefont {Cornell}},\
  }\bibfield  {title} {\bibinfo {title} {Vortices in a bose-einstein
  condensate},\ }\href {https://doi.org/10.1103/PhysRevLett.83.2498} {\bibfield
   {journal} {\bibinfo  {journal} {Phys. Rev. Lett.}\ }\textbf {\bibinfo
  {volume} {83}},\ \bibinfo {pages} {2498} (\bibinfo {year}
  {1999})}\BibitemShut {NoStop}%
\bibitem [{\citenamefont {Madison}\ \emph {et~al.}(2000)\citenamefont
  {Madison}, \citenamefont {Chevy}, \citenamefont {Wohlleben},\ and\
  \citenamefont {Dalibard}}]{PhysRevLett.84.806}%
  \BibitemOpen
  \bibfield  {author} {\bibinfo {author} {\bibfnamefont {K.~W.}\ \bibnamefont
  {Madison}}, \bibinfo {author} {\bibfnamefont {F.}~\bibnamefont {Chevy}},
  \bibinfo {author} {\bibfnamefont {W.}~\bibnamefont {Wohlleben}},\ and\
  \bibinfo {author} {\bibfnamefont {J.}~\bibnamefont {Dalibard}},\ }\bibfield
  {title} {\bibinfo {title} {Vortex formation in a stirred bose-einstein
  condensate},\ }\href {https://doi.org/10.1103/PhysRevLett.84.806} {\bibfield
  {journal} {\bibinfo  {journal} {Phys. Rev. Lett.}\ }\textbf {\bibinfo
  {volume} {84}},\ \bibinfo {pages} {806} (\bibinfo {year} {2000})}\BibitemShut
  {NoStop}%
\bibitem [{\citenamefont {Inouye}\ \emph {et~al.}(2001)\citenamefont {Inouye},
  \citenamefont {Gupta}, \citenamefont {Rosenband}, \citenamefont {Chikkatur},
  \citenamefont {G\"orlitz}, \citenamefont {Gustavson}, \citenamefont
  {Leanhardt}, \citenamefont {Pritchard},\ and\ \citenamefont
  {Ketterle}}]{PhysRevLett.87.080402}%
  \BibitemOpen
  \bibfield  {author} {\bibinfo {author} {\bibfnamefont {S.}~\bibnamefont
  {Inouye}}, \bibinfo {author} {\bibfnamefont {S.}~\bibnamefont {Gupta}},
  \bibinfo {author} {\bibfnamefont {T.}~\bibnamefont {Rosenband}}, \bibinfo
  {author} {\bibfnamefont {A.~P.}\ \bibnamefont {Chikkatur}}, \bibinfo {author}
  {\bibfnamefont {A.}~\bibnamefont {G\"orlitz}}, \bibinfo {author}
  {\bibfnamefont {T.~L.}\ \bibnamefont {Gustavson}}, \bibinfo {author}
  {\bibfnamefont {A.~E.}\ \bibnamefont {Leanhardt}}, \bibinfo {author}
  {\bibfnamefont {D.~E.}\ \bibnamefont {Pritchard}},\ and\ \bibinfo {author}
  {\bibfnamefont {W.}~\bibnamefont {Ketterle}},\ }\bibfield  {title} {\bibinfo
  {title} {Observation of vortex phase singularities in bose-einstein
  condensates},\ }\href {https://doi.org/10.1103/PhysRevLett.87.080402}
  {\bibfield  {journal} {\bibinfo  {journal} {Phys. Rev. Lett.}\ }\textbf
  {\bibinfo {volume} {87}},\ \bibinfo {pages} {080402} (\bibinfo {year}
  {2001})}\BibitemShut {NoStop}%
\bibitem [{\citenamefont {Beckmann}\ \emph {et~al.}(1979)\citenamefont
  {Beckmann}, \citenamefont {Karsch},\ and\ \citenamefont
  {Miller}}]{PhysRevLett.43.1277}%
  \BibitemOpen
  \bibfield  {author} {\bibinfo {author} {\bibfnamefont {R.}~\bibnamefont
  {Beckmann}}, \bibinfo {author} {\bibfnamefont {F.}~\bibnamefont {Karsch}},\
  and\ \bibinfo {author} {\bibfnamefont {D.~E.}\ \bibnamefont {Miller}},\
  }\bibfield  {title} {\bibinfo {title} {Bose-einstein condensation of a
  relativistic gas in $d$ dimensions},\ }\href
  {https://doi.org/10.1103/PhysRevLett.43.1277} {\bibfield  {journal} {\bibinfo
   {journal} {Phys. Rev. Lett.}\ }\textbf {\bibinfo {volume} {43}},\ \bibinfo
  {pages} {1277} (\bibinfo {year} {1979})}\BibitemShut {NoStop}%
\bibitem [{\citenamefont {Grether}\ \emph {et~al.}(2007)\citenamefont
  {Grether}, \citenamefont {de~Llano},\ and\ \citenamefont
  {Baker}}]{PhysRevLett.99.200406}%
  \BibitemOpen
  \bibfield  {author} {\bibinfo {author} {\bibfnamefont {M.}~\bibnamefont
  {Grether}}, \bibinfo {author} {\bibfnamefont {M.}~\bibnamefont {de~Llano}},\
  and\ \bibinfo {author} {\bibfnamefont {G.~A.}\ \bibnamefont {Baker}},\
  }\bibfield  {title} {\bibinfo {title} {Bose-einstein condensation in the
  relativistic ideal bose gas},\ }\href
  {https://doi.org/10.1103/PhysRevLett.99.200406} {\bibfield  {journal}
  {\bibinfo  {journal} {Phys. Rev. Lett.}\ }\textbf {\bibinfo {volume} {99}},\
  \bibinfo {pages} {200406} (\bibinfo {year} {2007})}\BibitemShut {NoStop}%
\bibitem [{\citenamefont {Fagnocchi}\ \emph {et~al.}(2010)\citenamefont
  {Fagnocchi}, \citenamefont {Finazzi}, \citenamefont {Liberati}, \citenamefont
  {Kormos},\ and\ \citenamefont {Trombettoni}}]{Fagnocchi_2010}%
  \BibitemOpen
  \bibfield  {author} {\bibinfo {author} {\bibfnamefont {S.}~\bibnamefont
  {Fagnocchi}}, \bibinfo {author} {\bibfnamefont {S.}~\bibnamefont {Finazzi}},
  \bibinfo {author} {\bibfnamefont {S.}~\bibnamefont {Liberati}}, \bibinfo
  {author} {\bibfnamefont {M.}~\bibnamefont {Kormos}},\ and\ \bibinfo {author}
  {\bibfnamefont {A.}~\bibnamefont {Trombettoni}},\ }\bibfield  {title}
  {\bibinfo {title} {Relativistic bose{\textendash}einstein condensates: a new
  system for analogue models of gravity},\ }\href
  {https://doi.org/10.1088/1367-2630/12/9/095012} {\bibfield  {journal}
  {\bibinfo  {journal} {New Journal of Physics}\ }\textbf {\bibinfo {volume}
  {12}},\ \bibinfo {pages} {095012} (\bibinfo {year} {2010})}\BibitemShut
  {NoStop}%
\bibitem [{\citenamefont {Cook}\ \emph {et~al.}(1994)\citenamefont {Cook},
  \citenamefont {Shapiro},\ and\ \citenamefont {Teukolsky}}]{Cook:1993qr}%
  \BibitemOpen
  \bibfield  {author} {\bibinfo {author} {\bibfnamefont {G.~B.}\ \bibnamefont
  {Cook}}, \bibinfo {author} {\bibfnamefont {S.~L.}\ \bibnamefont {Shapiro}},\
  and\ \bibinfo {author} {\bibfnamefont {S.~A.}\ \bibnamefont {Teukolsky}},\
  }\bibfield  {title} {\bibinfo {title} {{Rapidly rotating neutron stars in
  general relativity: Realistic equations of state}},\ }\href
  {https://doi.org/10.1086/173934} {\bibfield  {journal} {\bibinfo  {journal}
  {Astrophys. J.}\ }\textbf {\bibinfo {volume} {424}},\ \bibinfo {pages} {823}
  (\bibinfo {year} {1994})}\BibitemShut {NoStop}%
\bibitem [{\citenamefont {Bocquet}\ \emph {et~al.}(1995)\citenamefont
  {Bocquet}, \citenamefont {Bonazzola}, \citenamefont {Gourgoulhon},\ and\
  \citenamefont {Novak}}]{Bocquet:1995je}%
  \BibitemOpen
  \bibfield  {author} {\bibinfo {author} {\bibfnamefont {M.}~\bibnamefont
  {Bocquet}}, \bibinfo {author} {\bibfnamefont {S.}~\bibnamefont {Bonazzola}},
  \bibinfo {author} {\bibfnamefont {E.}~\bibnamefont {Gourgoulhon}},\ and\
  \bibinfo {author} {\bibfnamefont {J.}~\bibnamefont {Novak}},\ }\bibfield
  {title} {\bibinfo {title} {{Rotating neutron star models with magnetic
  field}},\ }\href@noop {} {\bibfield  {journal} {\bibinfo  {journal} {Astron.
  Astrophys.}\ }\textbf {\bibinfo {volume} {301}},\ \bibinfo {pages} {757}
  (\bibinfo {year} {1995})}\BibitemShut {NoStop}%
\bibitem [{\citenamefont {Andersson}(1998)}]{Andersson_1998}%
  \BibitemOpen
  \bibfield  {author} {\bibinfo {author} {\bibfnamefont {N.}~\bibnamefont
  {Andersson}},\ }\bibfield  {title} {\bibinfo {title} {A new class of unstable
  modes of rotating relativistic stars},\ }\href
  {https://doi.org/10.1086/305919} {\bibfield  {journal} {\bibinfo  {journal}
  {The Astrophysical Journal}\ }\textbf {\bibinfo {volume} {502}},\ \bibinfo
  {pages} {708} (\bibinfo {year} {1998})}\BibitemShut {NoStop}%
\bibitem [{\citenamefont {Marronetti}\ \emph {et~al.}(2008)\citenamefont
  {Marronetti}, \citenamefont {Tichy}, \citenamefont {Bruegmann}, \citenamefont
  {Gonzalez},\ and\ \citenamefont {Sperhake}}]{Marronetti:2007wz}%
  \BibitemOpen
  \bibfield  {author} {\bibinfo {author} {\bibfnamefont {P.}~\bibnamefont
  {Marronetti}}, \bibinfo {author} {\bibfnamefont {W.}~\bibnamefont {Tichy}},
  \bibinfo {author} {\bibfnamefont {B.}~\bibnamefont {Bruegmann}}, \bibinfo
  {author} {\bibfnamefont {J.}~\bibnamefont {Gonzalez}},\ and\ \bibinfo
  {author} {\bibfnamefont {U.}~\bibnamefont {Sperhake}},\ }\bibfield  {title}
  {\bibinfo {title} {{High-spin binary black hole mergers}},\ }\href
  {https://doi.org/10.1103/PhysRevD.77.064010} {\bibfield  {journal} {\bibinfo
  {journal} {Phys. Rev. D}\ }\textbf {\bibinfo {volume} {77}},\ \bibinfo
  {pages} {064010} (\bibinfo {year} {2008})}\BibitemShut {NoStop}%
\bibitem [{\citenamefont {Pook-Kolb}\ \emph {et~al.}(2019)\citenamefont
  {Pook-Kolb}, \citenamefont {Birnholtz}, \citenamefont {Krishnan},\ and\
  \citenamefont {Schnetter}}]{Pook-Kolb:2019iao}%
  \BibitemOpen
  \bibfield  {author} {\bibinfo {author} {\bibfnamefont {D.}~\bibnamefont
  {Pook-Kolb}}, \bibinfo {author} {\bibfnamefont {O.}~\bibnamefont
  {Birnholtz}}, \bibinfo {author} {\bibfnamefont {B.}~\bibnamefont
  {Krishnan}},\ and\ \bibinfo {author} {\bibfnamefont {E.}~\bibnamefont
  {Schnetter}},\ }\bibfield  {title} {\bibinfo {title} {{Interior of a Binary
  Black Hole Merger}},\ }\href {https://doi.org/10.1103/PhysRevLett.123.171102}
  {\bibfield  {journal} {\bibinfo  {journal} {Phys. Rev. Lett.}\ }\textbf
  {\bibinfo {volume} {123}},\ \bibinfo {pages} {171102} (\bibinfo {year}
  {2019})}\BibitemShut {NoStop}%
\bibitem [{\citenamefont {Becattini}\ \emph {et~al.}(2008)\citenamefont
  {Becattini}, \citenamefont {Piccinini},\ and\ \citenamefont
  {Rizzo}}]{Becattini:2007sr}%
  \BibitemOpen
  \bibfield  {author} {\bibinfo {author} {\bibfnamefont {F.}~\bibnamefont
  {Becattini}}, \bibinfo {author} {\bibfnamefont {F.}~\bibnamefont
  {Piccinini}},\ and\ \bibinfo {author} {\bibfnamefont {J.}~\bibnamefont
  {Rizzo}},\ }\bibfield  {title} {\bibinfo {title} {{Angular momentum
  conservation in heavy ion collisions at very high energy}},\ }\href
  {https://doi.org/10.1103/PhysRevC.77.024906} {\bibfield  {journal} {\bibinfo
  {journal} {Phys. Rev. C}\ }\textbf {\bibinfo {volume} {77}},\ \bibinfo
  {pages} {024906} (\bibinfo {year} {2008})}\BibitemShut {NoStop}%
\bibitem [{\citenamefont {Becattini}\ \emph {et~al.}(2015)\citenamefont
  {Becattini}, \citenamefont {Inghirami}, \citenamefont {Rolando},
  \citenamefont {Beraudo}, \citenamefont {Del~Zanna}, \citenamefont {De~Pace},
  \citenamefont {Nardi}, \citenamefont {Pagliara},\ and\ \citenamefont
  {Chandra}}]{Becattini:2015ska}%
  \BibitemOpen
  \bibfield  {author} {\bibinfo {author} {\bibfnamefont {F.}~\bibnamefont
  {Becattini}}, \bibinfo {author} {\bibfnamefont {G.}~\bibnamefont
  {Inghirami}}, \bibinfo {author} {\bibfnamefont {V.}~\bibnamefont {Rolando}},
  \bibinfo {author} {\bibfnamefont {A.}~\bibnamefont {Beraudo}}, \bibinfo
  {author} {\bibfnamefont {L.}~\bibnamefont {Del~Zanna}}, \bibinfo {author}
  {\bibfnamefont {A.}~\bibnamefont {De~Pace}}, \bibinfo {author} {\bibfnamefont
  {M.}~\bibnamefont {Nardi}}, \bibinfo {author} {\bibfnamefont
  {G.}~\bibnamefont {Pagliara}},\ and\ \bibinfo {author} {\bibfnamefont
  {V.}~\bibnamefont {Chandra}},\ }\bibfield  {title} {\bibinfo {title} {{A
  study of vorticity formation in high energy nuclear collisions}},\ }\href
  {https://doi.org/10.1140/epjc/s10052-015-3624-1} {\bibfield  {journal}
  {\bibinfo  {journal} {Eur. Phys. J. C}\ }\textbf {\bibinfo {volume} {75}},\
  \bibinfo {pages} {406} (\bibinfo {year} {2015})},\ \bibinfo {note} {[Erratum:
  Eur.Phys.J.C 78, 354 (2018)]}\BibitemShut {NoStop}%
\bibitem [{\citenamefont {Jiang}\ \emph {et~al.}(2016)\citenamefont {Jiang},
  \citenamefont {Lin},\ and\ \citenamefont {Liao}}]{Jiang:2016woz}%
  \BibitemOpen
  \bibfield  {author} {\bibinfo {author} {\bibfnamefont {Y.}~\bibnamefont
  {Jiang}}, \bibinfo {author} {\bibfnamefont {Z.}~\bibnamefont {Lin}},\ and\
  \bibinfo {author} {\bibfnamefont {J.}~\bibnamefont {Liao}},\ }\bibfield
  {title} {\bibinfo {title} {{Rotating quark-gluon plasma in relativistic heavy
  ion collisions}},\ }\href {https://doi.org/10.1103/PhysRevC.94.044910}
  {\bibfield  {journal} {\bibinfo  {journal} {Phys. Rev. C}\ }\textbf {\bibinfo
  {volume} {94}},\ \bibinfo {pages} {044910} (\bibinfo {year} {2016})},\
  \bibinfo {note} {[Erratum: Phys.Rev.C 95, 049904 (2017)]}\BibitemShut
  {NoStop}%
\bibitem [{\citenamefont {Voronyuk}\ \emph {et~al.}(2011)\citenamefont
  {Voronyuk}, \citenamefont {Toneev}, \citenamefont {Cassing}, \citenamefont
  {Bratkovskaya}, \citenamefont {Konchakovski},\ and\ \citenamefont
  {Voloshin}}]{Voronyuk:2011jd}%
  \BibitemOpen
  \bibfield  {author} {\bibinfo {author} {\bibfnamefont {V.}~\bibnamefont
  {Voronyuk}}, \bibinfo {author} {\bibfnamefont {V.~D.}\ \bibnamefont
  {Toneev}}, \bibinfo {author} {\bibfnamefont {W.}~\bibnamefont {Cassing}},
  \bibinfo {author} {\bibfnamefont {E.~L.}\ \bibnamefont {Bratkovskaya}},
  \bibinfo {author} {\bibfnamefont {V.~P.}\ \bibnamefont {Konchakovski}},\ and\
  \bibinfo {author} {\bibfnamefont {S.~A.}\ \bibnamefont {Voloshin}},\
  }\bibfield  {title} {\bibinfo {title} {{(Electro-)Magnetic field evolution in
  relativistic heavy-ion collisions}},\ }\href
  {https://doi.org/10.1103/PhysRevC.83.054911} {\bibfield  {journal} {\bibinfo
  {journal} {Phys. Rev. C}\ }\textbf {\bibinfo {volume} {83}},\ \bibinfo
  {pages} {054911} (\bibinfo {year} {2011})}\BibitemShut {NoStop}%
\bibitem [{\citenamefont {Bzdak}\ and\ \citenamefont
  {Skokov}(2012)}]{Bzdak:2011yy}%
  \BibitemOpen
  \bibfield  {author} {\bibinfo {author} {\bibfnamefont {A.}~\bibnamefont
  {Bzdak}}\ and\ \bibinfo {author} {\bibfnamefont {V.}~\bibnamefont {Skokov}},\
  }\bibfield  {title} {\bibinfo {title} {{Event-by-event fluctuations of
  magnetic and electric fields in heavy ion collisions}},\ }\href
  {https://doi.org/10.1016/j.physletb.2012.02.065} {\bibfield  {journal}
  {\bibinfo  {journal} {Phys. Lett. B}\ }\textbf {\bibinfo {volume} {710}},\
  \bibinfo {pages} {171} (\bibinfo {year} {2012})}\BibitemShut {NoStop}%
\bibitem [{\citenamefont {Kharzeev}\ \emph {et~al.}(2016)\citenamefont
  {Kharzeev}, \citenamefont {Liao}, \citenamefont {Voloshin},\ and\
  \citenamefont {Wang}}]{Kharzeev:2015znc}%
  \BibitemOpen
  \bibfield  {author} {\bibinfo {author} {\bibfnamefont {D.~E.}\ \bibnamefont
  {Kharzeev}}, \bibinfo {author} {\bibfnamefont {J.}~\bibnamefont {Liao}},
  \bibinfo {author} {\bibfnamefont {S.~A.}\ \bibnamefont {Voloshin}},\ and\
  \bibinfo {author} {\bibfnamefont {G.}~\bibnamefont {Wang}},\ }\bibfield
  {title} {\bibinfo {title} {{Chiral magnetic and vortical effects in
  high-energy nuclear collisions\textemdash{}A status report}},\ }\href
  {https://doi.org/10.1016/j.ppnp.2016.01.001} {\bibfield  {journal} {\bibinfo
  {journal} {Prog. Part. Nucl. Phys.}\ }\textbf {\bibinfo {volume} {88}},\
  \bibinfo {pages} {1} (\bibinfo {year} {2016})}\BibitemShut {NoStop}%
\bibitem [{\citenamefont {Tuchin}(2013)}]{Tuchin:2013ie}%
  \BibitemOpen
  \bibfield  {author} {\bibinfo {author} {\bibfnamefont {K.}~\bibnamefont
  {Tuchin}},\ }\bibfield  {title} {\bibinfo {title} {{Particle production in
  strong electromagnetic fields in relativistic heavy-ion collisions}},\ }\href
  {https://doi.org/10.1155/2013/490495} {\bibfield  {journal} {\bibinfo
  {journal} {Adv. High Energy Phys.}\ }\textbf {\bibinfo {volume} {2013}},\
  \bibinfo {pages} {490495} (\bibinfo {year} {2013})}\BibitemShut {NoStop}%
\bibitem [{\citenamefont {Abdallah}\ \emph {et~al.}(2022)\citenamefont
  {Abdallah} \emph {et~al.}}]{STAR:2021mii}%
  \BibitemOpen
  \bibfield  {author} {\bibinfo {author} {\bibfnamefont {M.}~\bibnamefont
  {Abdallah}} \emph {et~al.} (\bibinfo {collaboration} {STAR}),\ }\bibfield
  {title} {\bibinfo {title} {{Search for the chiral magnetic effect with isobar
  collisions at $\sqrt {s_{NN}}$=200 GeV by the STAR Collaboration at the BNL
  Relativistic Heavy Ion Collider}},\ }\href
  {https://doi.org/10.1103/PhysRevC.105.014901} {\bibfield  {journal} {\bibinfo
   {journal} {Phys. Rev. C}\ }\textbf {\bibinfo {volume} {105}},\ \bibinfo
  {pages} {014901} (\bibinfo {year} {2022})}\BibitemShut {NoStop}%
\bibitem [{\citenamefont {Deng}\ and\ \citenamefont
  {Huang}(2016)}]{Deng:2016gyh}%
  \BibitemOpen
  \bibfield  {author} {\bibinfo {author} {\bibfnamefont {W.}~\bibnamefont
  {Deng}}\ and\ \bibinfo {author} {\bibfnamefont {X.}~\bibnamefont {Huang}},\
  }\bibfield  {title} {\bibinfo {title} {{Vorticity in Heavy-Ion Collisions}},\
  }\href {https://doi.org/10.1103/PhysRevC.93.064907} {\bibfield  {journal}
  {\bibinfo  {journal} {Phys. Rev. C}\ }\textbf {\bibinfo {volume} {93}},\
  \bibinfo {pages} {064907} (\bibinfo {year} {2016})}\BibitemShut {NoStop}%
\bibitem [{\citenamefont {Becattini}\ and\ \citenamefont
  {Lisa}(2020)}]{Becattini:2020ngo}%
  \BibitemOpen
  \bibfield  {author} {\bibinfo {author} {\bibfnamefont {F.}~\bibnamefont
  {Becattini}}\ and\ \bibinfo {author} {\bibfnamefont {M.~A.}\ \bibnamefont
  {Lisa}},\ }\bibfield  {title} {\bibinfo {title} {{Polarization and Vorticity
  in the Quark\textendash{}Gluon Plasma}},\ }\href
  {https://doi.org/10.1146/annurev-nucl-021920-095245} {\bibfield  {journal}
  {\bibinfo  {journal} {Ann. Rev. Nucl. Part. Sci.}\ }\textbf {\bibinfo
  {volume} {70}},\ \bibinfo {pages} {395} (\bibinfo {year} {2020})}\BibitemShut
  {NoStop}%
\bibitem [{\citenamefont {Adamczyk}\ \emph {et~al.}(2017)\citenamefont
  {Adamczyk} \emph {et~al.}}]{STAR:2017ckg}%
  \BibitemOpen
  \bibfield  {author} {\bibinfo {author} {\bibfnamefont {L.}~\bibnamefont
  {Adamczyk}} \emph {et~al.} (\bibinfo {collaboration} {STAR}),\ }\bibfield
  {title} {\bibinfo {title} {{Global $\Lambda$ hyperon polarization in nuclear
  collisions: evidence for the most vortical fluid}},\ }\href
  {https://doi.org/10.1038/nature23004} {\bibfield  {journal} {\bibinfo
  {journal} {Nature}\ }\textbf {\bibinfo {volume} {548}},\ \bibinfo {pages}
  {62} (\bibinfo {year} {2017})}\BibitemShut {NoStop}%
\bibitem [{\citenamefont {Adam}\ \emph {et~al.}(2019)\citenamefont {Adam} \emph
  {et~al.}}]{STAR:2019erd}%
  \BibitemOpen
  \bibfield  {author} {\bibinfo {author} {\bibfnamefont {J.}~\bibnamefont
  {Adam}} \emph {et~al.} (\bibinfo {collaboration} {STAR}),\ }\bibfield
  {title} {\bibinfo {title} {{Polarization of $\Lambda$ ($\bar{\Lambda}$)
  hyperons along the beam direction in Au+Au collisions at $\sqrt{s_{_{NN}}}$ =
  200 GeV}},\ }\href {https://doi.org/10.1103/PhysRevLett.123.132301}
  {\bibfield  {journal} {\bibinfo  {journal} {Phys. Rev. Lett.}\ }\textbf
  {\bibinfo {volume} {123}},\ \bibinfo {pages} {132301} (\bibinfo {year}
  {2019})}\BibitemShut {NoStop}%
\bibitem [{\citenamefont {Klimenko}(1992)}]{Klimenko:1991he}%
  \BibitemOpen
  \bibfield  {author} {\bibinfo {author} {\bibfnamefont {K.~G.}\ \bibnamefont
  {Klimenko}},\ }\bibfield  {title} {\bibinfo {title} {{Three-dimensional
  Gross-Neveu model at nonzero temperature and in an external magnetic
  field}},\ }\href {https://doi.org/10.1007/BF01566663} {\bibfield  {journal}
  {\bibinfo  {journal} {Z. Phys. C}\ }\textbf {\bibinfo {volume} {54}},\
  \bibinfo {pages} {323} (\bibinfo {year} {1992})}\BibitemShut {NoStop}%
\bibitem [{\citenamefont {Gusynin}\ \emph {et~al.}(1994)\citenamefont
  {Gusynin}, \citenamefont {Miransky},\ and\ \citenamefont
  {Shovkovy}}]{PhysRevLett.73.3499}%
  \BibitemOpen
  \bibfield  {author} {\bibinfo {author} {\bibfnamefont {V.~P.}\ \bibnamefont
  {Gusynin}}, \bibinfo {author} {\bibfnamefont {V.~A.}\ \bibnamefont
  {Miransky}},\ and\ \bibinfo {author} {\bibfnamefont {I.~A.}\ \bibnamefont
  {Shovkovy}},\ }\bibfield  {title} {\bibinfo {title} {Catalysis of dynamical
  flavor symmetry breaking by a magnetic field in 2 + 1 dimensions},\ }\href
  {https://doi.org/10.1103/PhysRevLett.73.3499} {\bibfield  {journal} {\bibinfo
   {journal} {Phys. Rev. Lett.}\ }\textbf {\bibinfo {volume} {73}},\ \bibinfo
  {pages} {3499} (\bibinfo {year} {1994})}\BibitemShut {NoStop}%
\bibitem [{\citenamefont {Miransky}\ and\ \citenamefont
  {Shovkovy}(2015)}]{Miransky:2015ava}%
  \BibitemOpen
  \bibfield  {author} {\bibinfo {author} {\bibfnamefont {V.~A.}\ \bibnamefont
  {Miransky}}\ and\ \bibinfo {author} {\bibfnamefont {I.~A.}\ \bibnamefont
  {Shovkovy}},\ }\bibfield  {title} {\bibinfo {title} {{Quantum field theory in
  a magnetic field: From quantum chromodynamics to graphene and Dirac
  semimetals}},\ }\href {https://doi.org/10.1016/j.physrep.2015.02.003}
  {\bibfield  {journal} {\bibinfo  {journal} {Phys. Rept.}\ }\textbf {\bibinfo
  {volume} {576}},\ \bibinfo {pages} {1} (\bibinfo {year} {2015})}\BibitemShut
  {NoStop}%
\bibitem [{\citenamefont {Kharzeev}\ and\ \citenamefont
  {Zhitnitsky}(2007)}]{Kharzeev:2007tn}%
  \BibitemOpen
  \bibfield  {author} {\bibinfo {author} {\bibfnamefont {D.}~\bibnamefont
  {Kharzeev}}\ and\ \bibinfo {author} {\bibfnamefont {A.}~\bibnamefont
  {Zhitnitsky}},\ }\bibfield  {title} {\bibinfo {title} {{Charge separation
  induced by P-odd bubbles in QCD matter}},\ }\href
  {https://doi.org/10.1016/j.nuclphysa.2007.10.001} {\bibfield  {journal}
  {\bibinfo  {journal} {Nucl. Phys. A}\ }\textbf {\bibinfo {volume} {797}},\
  \bibinfo {pages} {67} (\bibinfo {year} {2007})}\BibitemShut {NoStop}%
\bibitem [{\citenamefont {Kharzeev}\ and\ \citenamefont
  {Son}(2011)}]{Kharzeev:2010gr}%
  \BibitemOpen
  \bibfield  {author} {\bibinfo {author} {\bibfnamefont {D.~E.}\ \bibnamefont
  {Kharzeev}}\ and\ \bibinfo {author} {\bibfnamefont {D.~T.}\ \bibnamefont
  {Son}},\ }\bibfield  {title} {\bibinfo {title} {{Testing the chiral magnetic
  and chiral vortical effects in heavy ion collisions}},\ }\href
  {https://doi.org/10.1103/PhysRevLett.106.062301} {\bibfield  {journal}
  {\bibinfo  {journal} {Phys. Rev. Lett.}\ }\textbf {\bibinfo {volume} {106}},\
  \bibinfo {pages} {062301} (\bibinfo {year} {2011})}\BibitemShut {NoStop}%
\bibitem [{\citenamefont {Jiang}\ \emph {et~al.}(2015)\citenamefont {Jiang},
  \citenamefont {Huang},\ and\ \citenamefont {Liao}}]{Jiang:2015cva}%
  \BibitemOpen
  \bibfield  {author} {\bibinfo {author} {\bibfnamefont {Y.}~\bibnamefont
  {Jiang}}, \bibinfo {author} {\bibfnamefont {X.-G.}\ \bibnamefont {Huang}},\
  and\ \bibinfo {author} {\bibfnamefont {J.}~\bibnamefont {Liao}},\ }\bibfield
  {title} {\bibinfo {title} {{Chiral vortical wave and induced flavor charge
  transport in a rotating quark-gluon plasma}},\ }\href
  {https://doi.org/10.1103/PhysRevD.92.071501} {\bibfield  {journal} {\bibinfo
  {journal} {Phys. Rev. D}\ }\textbf {\bibinfo {volume} {92}},\ \bibinfo
  {pages} {071501} (\bibinfo {year} {2015})}\BibitemShut {NoStop}%
\bibitem [{\citenamefont {Kamada}\ \emph {et~al.}(2023)\citenamefont {Kamada},
  \citenamefont {Yamamoto},\ and\ \citenamefont {Yang}}]{Kamada:2022nyt}%
  \BibitemOpen
  \bibfield  {author} {\bibinfo {author} {\bibfnamefont {K.}~\bibnamefont
  {Kamada}}, \bibinfo {author} {\bibfnamefont {N.}~\bibnamefont {Yamamoto}},\
  and\ \bibinfo {author} {\bibfnamefont {D.-L.}\ \bibnamefont {Yang}},\
  }\bibfield  {title} {\bibinfo {title} {{Chiral effects in astrophysics and
  cosmology}},\ }\href {https://doi.org/10.1016/j.ppnp.2022.104016} {\bibfield
  {journal} {\bibinfo  {journal} {Prog. Part. Nucl. Phys.}\ }\textbf {\bibinfo
  {volume} {129}},\ \bibinfo {pages} {104016} (\bibinfo {year}
  {2023})}\BibitemShut {NoStop}%
\bibitem [{\citenamefont {Jiang}\ and\ \citenamefont
  {Liao}(2016)}]{Jiang:2016wvv}%
  \BibitemOpen
  \bibfield  {author} {\bibinfo {author} {\bibfnamefont {Y.}~\bibnamefont
  {Jiang}}\ and\ \bibinfo {author} {\bibfnamefont {J.}~\bibnamefont {Liao}},\
  }\bibfield  {title} {\bibinfo {title} {{Pairing Phase Transitions of Matter
  under Rotation}},\ }\href {https://doi.org/10.1103/PhysRevLett.117.192302}
  {\bibfield  {journal} {\bibinfo  {journal} {Phys. Rev. Lett.}\ }\textbf
  {\bibinfo {volume} {117}},\ \bibinfo {pages} {192302} (\bibinfo {year}
  {2016})}\BibitemShut {NoStop}%
\bibitem [{\citenamefont {Ebihara}\ \emph {et~al.}(2017)\citenamefont
  {Ebihara}, \citenamefont {Fukushima},\ and\ \citenamefont
  {Mameda}}]{Ebihara:2016fwa}%
  \BibitemOpen
  \bibfield  {author} {\bibinfo {author} {\bibfnamefont {S.}~\bibnamefont
  {Ebihara}}, \bibinfo {author} {\bibfnamefont {K.}~\bibnamefont {Fukushima}},\
  and\ \bibinfo {author} {\bibfnamefont {K.}~\bibnamefont {Mameda}},\
  }\bibfield  {title} {\bibinfo {title} {{Boundary effects and gapped
  dispersion in rotating fermionic matter}},\ }\href
  {https://doi.org/10.1016/j.physletb.2016.11.010} {\bibfield  {journal}
  {\bibinfo  {journal} {Phys. Lett. B}\ }\textbf {\bibinfo {volume} {764}},\
  \bibinfo {pages} {94} (\bibinfo {year} {2017})}\BibitemShut {NoStop}%
\bibitem [{\citenamefont {Chen}\ \emph {et~al.}(2022)\citenamefont {Chen},
  \citenamefont {Li},\ and\ \citenamefont {Huang}}]{Chen:2022mhf}%
  \BibitemOpen
  \bibfield  {author} {\bibinfo {author} {\bibfnamefont {Y.}~\bibnamefont
  {Chen}}, \bibinfo {author} {\bibfnamefont {D.}~\bibnamefont {Li}},\ and\
  \bibinfo {author} {\bibfnamefont {M.}~\bibnamefont {Huang}},\ }\bibfield
  {title} {\bibinfo {title} {{Inhomogeneous chiral condensation under rotation
  in the holographic QCD}},\ }\href
  {https://doi.org/10.1103/PhysRevD.106.106002} {\bibfield  {journal} {\bibinfo
   {journal} {Phys. Rev. D}\ }\textbf {\bibinfo {volume} {106}},\ \bibinfo
  {pages} {106002} (\bibinfo {year} {2022})}\BibitemShut {NoStop}%
\bibitem [{\citenamefont {Braguta}\ \emph {et~al.}(2021)\citenamefont
  {Braguta}, \citenamefont {Kotov}, \citenamefont {Kuznedelev},\ and\
  \citenamefont {Roenko}}]{Braguta:2021jgn}%
  \BibitemOpen
  \bibfield  {author} {\bibinfo {author} {\bibfnamefont {V.~V.}\ \bibnamefont
  {Braguta}}, \bibinfo {author} {\bibfnamefont {A.~Y.}\ \bibnamefont {Kotov}},
  \bibinfo {author} {\bibfnamefont {D.~D.}\ \bibnamefont {Kuznedelev}},\ and\
  \bibinfo {author} {\bibfnamefont {A.~A.}\ \bibnamefont {Roenko}},\ }\bibfield
   {title} {\bibinfo {title} {{Influence of relativistic rotation on the
  confinement-deconfinement transition in gluodynamics}},\ }\href
  {https://doi.org/10.1103/PhysRevD.103.094515} {\bibfield  {journal} {\bibinfo
   {journal} {Phys. Rev. D}\ }\textbf {\bibinfo {volume} {103}},\ \bibinfo
  {pages} {094515} (\bibinfo {year} {2021})}\BibitemShut {NoStop}%
\bibitem [{\citenamefont {Yang}\ and\ \citenamefont
  {Huang}(2023)}]{Yang:2023vsw}%
  \BibitemOpen
  \bibfield  {author} {\bibinfo {author} {\bibfnamefont {J.}~\bibnamefont
  {Yang}}\ and\ \bibinfo {author} {\bibfnamefont {X.}~\bibnamefont {Huang}},\
  }\bibfield  {title} {\bibinfo {title} {{QCD on Rotating Lattice with
  Staggered Fermions}},\ }\href@noop {} {\  (\bibinfo {year} {2023})},\ \Eprint
  {https://arxiv.org/abs/2307.05755} {arXiv:2307.05755 [hep-lat]} \BibitemShut
  {NoStop}%
\bibitem [{\citenamefont {Cao}(2024)}]{Cao:2023olg}%
  \BibitemOpen
  \bibfield  {author} {\bibinfo {author} {\bibfnamefont {G.}~\bibnamefont
  {Cao}},\ }\bibfield  {title} {\bibinfo {title} {{Effects of imaginary and
  real rotations on QCD matters}},\ }\href
  {https://doi.org/10.1103/PhysRevD.109.014001} {\bibfield  {journal} {\bibinfo
   {journal} {Phys. Rev. D}\ }\textbf {\bibinfo {volume} {109}},\ \bibinfo
  {pages} {014001} (\bibinfo {year} {2024})}\BibitemShut {NoStop}%
\bibitem [{\citenamefont {Iyer}(1982)}]{PhysRevD.26.1900}%
  \BibitemOpen
  \bibfield  {author} {\bibinfo {author} {\bibfnamefont {B.~R.}\ \bibnamefont
  {Iyer}},\ }\bibfield  {title} {\bibinfo {title} {Dirac field theory in
  rotating coordinates},\ }\href {https://doi.org/10.1103/PhysRevD.26.1900}
  {\bibfield  {journal} {\bibinfo  {journal} {Phys. Rev. D}\ }\textbf {\bibinfo
  {volume} {26}},\ \bibinfo {pages} {1900} (\bibinfo {year}
  {1982})}\BibitemShut {NoStop}%
\bibitem [{\citenamefont {Chen}\ \emph {et~al.}(2016)\citenamefont {Chen},
  \citenamefont {Fukushima}, \citenamefont {Huang},\ and\ \citenamefont
  {Mameda}}]{Chen:2015hfc}%
  \BibitemOpen
  \bibfield  {author} {\bibinfo {author} {\bibfnamefont {H.}~\bibnamefont
  {Chen}}, \bibinfo {author} {\bibfnamefont {K.}~\bibnamefont {Fukushima}},
  \bibinfo {author} {\bibfnamefont {X.}~\bibnamefont {Huang}},\ and\ \bibinfo
  {author} {\bibfnamefont {K.}~\bibnamefont {Mameda}},\ }\bibfield  {title}
  {\bibinfo {title} {{Analogy between rotation and density for Dirac fermions
  in a magnetic field}},\ }\href {https://doi.org/10.1103/PhysRevD.93.104052}
  {\bibfield  {journal} {\bibinfo  {journal} {Phys. Rev. D}\ }\textbf {\bibinfo
  {volume} {93}},\ \bibinfo {pages} {104052} (\bibinfo {year}
  {2016})}\BibitemShut {NoStop}%
\bibitem [{\citenamefont {Liu}\ and\ \citenamefont
  {Zahed}(2018{\natexlab{a}})}]{Liu:2017zhl}%
  \BibitemOpen
  \bibfield  {author} {\bibinfo {author} {\bibfnamefont {Y.}~\bibnamefont
  {Liu}}\ and\ \bibinfo {author} {\bibfnamefont {I.}~\bibnamefont {Zahed}},\
  }\bibfield  {title} {\bibinfo {title} {{Rotating Dirac fermions in a magnetic
  field in 1+2 and 1+3 dimensions}},\ }\href
  {https://doi.org/10.1103/PhysRevD.98.014017} {\bibfield  {journal} {\bibinfo
  {journal} {Phys. Rev. D}\ }\textbf {\bibinfo {volume} {98}},\ \bibinfo
  {pages} {014017} (\bibinfo {year} {2018}{\natexlab{a}})}\BibitemShut
  {NoStop}%
\bibitem [{\citenamefont {Fukushima}\ \emph {et~al.}(2020)\citenamefont
  {Fukushima}, \citenamefont {Shimazaki},\ and\ \citenamefont
  {Wang}}]{Fukushima:2020ncb}%
  \BibitemOpen
  \bibfield  {author} {\bibinfo {author} {\bibfnamefont {K.}~\bibnamefont
  {Fukushima}}, \bibinfo {author} {\bibfnamefont {T.}~\bibnamefont
  {Shimazaki}},\ and\ \bibinfo {author} {\bibfnamefont {L.}~\bibnamefont
  {Wang}},\ }\bibfield  {title} {\bibinfo {title} {{Mode decomposed chiral
  magnetic effect and rotating fermions}},\ }\href
  {https://doi.org/10.1103/PhysRevD.102.014045} {\bibfield  {journal} {\bibinfo
   {journal} {Phys. Rev. D}\ }\textbf {\bibinfo {volume} {102}},\ \bibinfo
  {pages} {014045} (\bibinfo {year} {2020})}\BibitemShut {NoStop}%
\bibitem [{\citenamefont {McInnes}(2016)}]{McInnes:2016dwk}%
  \BibitemOpen
  \bibfield  {author} {\bibinfo {author} {\bibfnamefont {B.}~\bibnamefont
  {McInnes}},\ }\bibfield  {title} {\bibinfo {title} {{A rotation/magnetism
  analogy for the quark\textendash{}gluon plasma}},\ }\href
  {https://doi.org/10.1016/j.nuclphysb.2016.08.001} {\bibfield  {journal}
  {\bibinfo  {journal} {Nucl. Phys. B}\ }\textbf {\bibinfo {volume} {911}},\
  \bibinfo {pages} {173} (\bibinfo {year} {2016})}\BibitemShut {NoStop}%
\bibitem [{\citenamefont {Hattori}\ and\ \citenamefont
  {Yin}(2016)}]{Hattori:2016njk}%
  \BibitemOpen
  \bibfield  {author} {\bibinfo {author} {\bibfnamefont {K.}~\bibnamefont
  {Hattori}}\ and\ \bibinfo {author} {\bibfnamefont {Y.}~\bibnamefont {Yin}},\
  }\bibfield  {title} {\bibinfo {title} {{Charge redistribution from anomalous
  magnetovorticity coupling}},\ }\href
  {https://doi.org/10.1103/PhysRevLett.117.152002} {\bibfield  {journal}
  {\bibinfo  {journal} {Phys. Rev. Lett.}\ }\textbf {\bibinfo {volume} {117}},\
  \bibinfo {pages} {152002} (\bibinfo {year} {2016})}\BibitemShut {NoStop}%
\bibitem [{\citenamefont {Chernodub}\ and\ \citenamefont
  {Gongyo}(2017)}]{Chernodub:2016kxh}%
  \BibitemOpen
  \bibfield  {author} {\bibinfo {author} {\bibfnamefont {M.~N.}\ \bibnamefont
  {Chernodub}}\ and\ \bibinfo {author} {\bibfnamefont {S.}~\bibnamefont
  {Gongyo}},\ }\bibfield  {title} {\bibinfo {title} {{Interacting fermions in
  rotation: chiral symmetry restoration, moment of inertia and
  thermodynamics}},\ }\href {https://doi.org/10.1007/JHEP01(2017)136}
  {\bibfield  {journal} {\bibinfo  {journal} {JHEP}\ }\textbf {\bibinfo
  {volume} {01}},\ \bibinfo {pages} {136}},\ \Eprint
  {https://arxiv.org/abs/1611.02598} {arXiv:1611.02598 [hep-th]} \BibitemShut
  {NoStop}%
\bibitem [{\citenamefont {Fukushima}(2019)}]{Fukushima:2018grm}%
  \BibitemOpen
  \bibfield  {author} {\bibinfo {author} {\bibfnamefont {K.}~\bibnamefont
  {Fukushima}},\ }\bibfield  {title} {\bibinfo {title} {{Extreme matter in
  electromagnetic fields and rotation}},\ }\href
  {https://doi.org/10.1016/j.ppnp.2019.04.001} {\bibfield  {journal} {\bibinfo
  {journal} {Prog. Part. Nucl. Phys.}\ }\textbf {\bibinfo {volume} {107}},\
  \bibinfo {pages} {167} (\bibinfo {year} {2019})}\BibitemShut {NoStop}%
\bibitem [{\citenamefont {Cao}(2021)}]{Cao:2020pmm}%
  \BibitemOpen
  \bibfield  {author} {\bibinfo {author} {\bibfnamefont {G.}~\bibnamefont
  {Cao}},\ }\bibfield  {title} {\bibinfo {title} {{Charged rho superconductor
  in the presence of magnetic field and rotation}},\ }\href
  {https://doi.org/10.1140/epjc/s10052-021-08900-8} {\bibfield  {journal}
  {\bibinfo  {journal} {Eur. Phys. J. C}\ }\textbf {\bibinfo {volume} {81}},\
  \bibinfo {pages} {148} (\bibinfo {year} {2021})}\BibitemShut {NoStop}%
\bibitem [{\citenamefont {Liu}\ and\ \citenamefont
  {Zahed}(2018{\natexlab{b}})}]{PhysRevLett.120.032001}%
  \BibitemOpen
  \bibfield  {author} {\bibinfo {author} {\bibfnamefont {Y.}~\bibnamefont
  {Liu}}\ and\ \bibinfo {author} {\bibfnamefont {I.}~\bibnamefont {Zahed}},\
  }\bibfield  {title} {\bibinfo {title} {Pion condensation by rotation in a
  magnetic field},\ }\href {https://doi.org/10.1103/PhysRevLett.120.032001}
  {\bibfield  {journal} {\bibinfo  {journal} {Phys. Rev. Lett.}\ }\textbf
  {\bibinfo {volume} {120}},\ \bibinfo {pages} {032001} (\bibinfo {year}
  {2018}{\natexlab{b}})}\BibitemShut {NoStop}%
\bibitem [{\citenamefont {Chen}\ \emph {et~al.}(2019)\citenamefont {Chen},
  \citenamefont {Huang},\ and\ \citenamefont {Mameda}}]{Chen:2019tcp}%
  \BibitemOpen
  \bibfield  {author} {\bibinfo {author} {\bibfnamefont {H.}~\bibnamefont
  {Chen}}, \bibinfo {author} {\bibfnamefont {X.}~\bibnamefont {Huang}},\ and\
  \bibinfo {author} {\bibfnamefont {K.}~\bibnamefont {Mameda}},\ }\href@noop {}
  {\bibinfo {title} {{Do charged pions condense in a magnetic field with
  rotation?}}} (\bibinfo {year} {2019}),\ \Eprint
  {https://arxiv.org/abs/1910.02700} {arXiv:1910.02700 [nucl-th]} \BibitemShut
  {NoStop}%
\bibitem [{\citenamefont {Guo}\ \emph {et~al.}(2022)\citenamefont {Guo},
  \citenamefont {Li}, \citenamefont {Mu},\ and\ \citenamefont
  {He}}]{Guo:2021gbz}%
  \BibitemOpen
  \bibfield  {author} {\bibinfo {author} {\bibfnamefont {T.}~\bibnamefont
  {Guo}}, \bibinfo {author} {\bibfnamefont {J.}~\bibnamefont {Li}}, \bibinfo
  {author} {\bibfnamefont {C.}~\bibnamefont {Mu}},\ and\ \bibinfo {author}
  {\bibfnamefont {L.}~\bibnamefont {He}},\ }\bibfield  {title} {\bibinfo
  {title} {{Formation of a supergiant quantum vortex in a relativistic
  Bose-Einstein condensate driven by rotation and a parallel magnetic field}},\
  }\href {https://doi.org/10.1103/PhysRevD.106.094010} {\bibfield  {journal}
  {\bibinfo  {journal} {Phys. Rev. D}\ }\textbf {\bibinfo {volume} {106}},\
  \bibinfo {pages} {094010} (\bibinfo {year} {2022})}\BibitemShut {NoStop}%
\bibitem [{\citenamefont {Cao}\ and\ \citenamefont {He}(2019)}]{Cao:2019ctl}%
  \BibitemOpen
  \bibfield  {author} {\bibinfo {author} {\bibfnamefont {G.}~\bibnamefont
  {Cao}}\ and\ \bibinfo {author} {\bibfnamefont {L.}~\bibnamefont {He}},\
  }\bibfield  {title} {\bibinfo {title} {{Rotation induced charged pion
  condensation in a strong magnetic field: A Nambu--Jona-Lasino model study}},\
  }\href {https://doi.org/10.1103/PhysRevD.100.094015} {\bibfield  {journal}
  {\bibinfo  {journal} {Phys. Rev. D}\ }\textbf {\bibinfo {volume} {100}},\
  \bibinfo {pages} {094015} (\bibinfo {year} {2019})}\BibitemShut {NoStop}%
\end{thebibliography}%

\end{document}